# Rubble Pile Asteroids

Kevin J. Walsh
Southwest Research Institute
1050 Walnut St. Suite 300
Boulder, CO 80302
kwalsh@boulder.swri.edu

**Abstract**: The moniker "rubble pile" is typically applied to all solar system bodies 200m < Diameter < ~10km - where in this size range there is an abundance of evidence that nearly every object is bound primarily by self-gravity with significant void space or bulk porosity between irregularly shaped constituent particles. The understanding of this population is derived from wide-ranging population studies of derived shape and spin, decades of observational studies in numerous wavelengths, evidence left behind from impacts on planets and moons and the *in situ* study of a few objects via spacecraft flyby or rendezvous. The internal structure, however, which is responsible for the name "rubble pile", is never directly observed, but belies a violent history. Many or most of the asteroids on near-Earth orbits, and the ones most accessible for rendezvous and *in situ* study, are likely byproducts of the continued collisional evolution of the Main Asteroid Belt.

# 1.0 "Rubble Pile"

There is no guarantee that a moniker, such as "rubble pile" will, once coined, hold up over time. Here, used as a term to qualitatively describe the bulk nature of small asteroids and possibly comets, the term dating ~40 years is still seemingly perfect (Chapman 1978; Chapman et al. 1978; Davis et al. 1985; Weissman 1986). In fact, the images of near-Earth asteroid Itokawa, returned by JAXA's Hayabusa space mission, could not have more clearly shown what scientists had inferred for decades about small asteroids. This small world looked like a pile of rocks from someone's garden or a mountain scree field. There are rough and angular boulders and cobbles littering the surface, one oddly large boulder looking out of place, and a few "ponds" of finer grains (Figure 1; Fujiwara 2006).

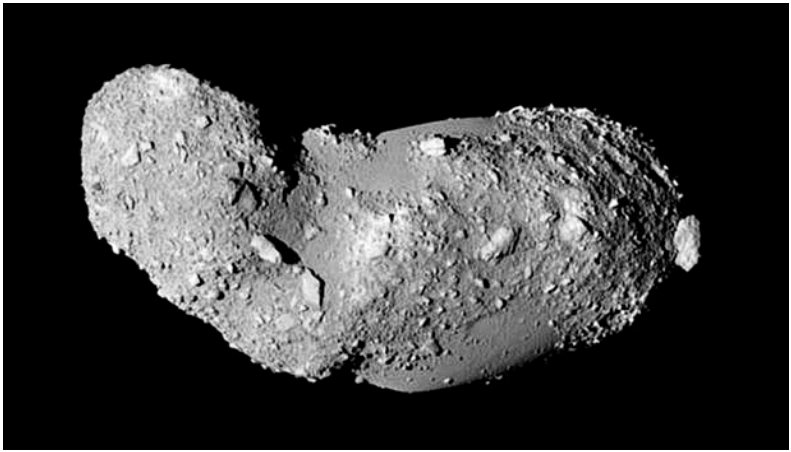

[**Figure 1:** A poster-child Rubble Pile, asteroid Itokawa. Image Credit & Copyright: ISAS, JAXA]

The term "rubble pile" pre-dates the images of Itokawa by nearly three decades and was thus built on evidence compiled from astronomical observations, theoretical models and numerous researchers' intuition. The definition of a rubble pile was codified in a parameter space of tensile strength and bulk porosity, to be bodies with zero, or nearly zero, tensile strength and moderate bulk porosity, and therefore significant internal void space (Richardson et al. 2002). They were defined to be different than objects that were shattered, and thus strengthless and bound by gravity alone, but organized with minimal void space or porosity - something that could be referred to as a "shattered aggregate". While this community-wide effort to clarify the properties of small bodies in Richardson et al. (2002) specified a name of "gravitational aggregate" for low-strength and high porosity objects, the terminology of "rubble pile" remained in common use, and fully returned to prominence with the publication of the spectacular images of Itokawa in an article titled: "The Rubble-Pile Asteroid Itokawa as Observed by Hayabusa" (Fujiwara et al. 2006). Thus we embrace this name in this work, and aim to describe the essence of a "rubble pile": an unorganized collection of macroscopic particles (rubble) held together by their self gravity.

Some of the most specific data used to build the characterization of a rubble pile come from the study of near-Earth asteroids since members of this population regularly have close passages by the Earth enabling a wide array of ground-based observations - including radar ranging and doppler measurements that can be used to map an object's shape, spin and some surface properties. The near-Earth objects (NEOs or NEAs for near-Earth asteroids) weren't formed on their current orbits and will only be around this part of the solar system for an average of ~10Myr before impacting a planet, the Sun or get ejected from the Solar System (Gladman et al. 2000). The population is transient and roughly in a steady-state, as asteroids that are removed due to collisions or ejection are replaced by new asteroids diffusing out of the Main Asteroid belt (Bottke et al. 2002b). Furthermore, the population contains all of the various spectral classes of asteroids found in the Main Belt, but they can't be directly linked back to a specific origin or orbit in the Main Belt. Significant modeling efforts provide only probabilistic accounting of their most likely pathways to reach their current orbits (Bottke et al. 2002b). Therefore we will largely skip a detailed discussion of the diverse taxonomy of asteroids found in the near-Earth and Main Belt populations, as they appear to have similar behavior in the data to be presented, despite having some different fundamental properties (amounts of volatiles, densities, spectra etc.: see DeMeo and Carry 2014). Similarly, most observed comets fit into the size range of typical rubble pile asteroids, and, by way of comet Shoemaker-Levy 9 and 67P/Churyumov–Gerasimenko, provided substantial interest and evidence for rubble pile structures. However, while comets will be mentioned both in passing the bulk of the evidence for, and data on, rubble piles will be asteroid focused.

The concept of solar system bodies having limited strength dates back to Jeffreys (1947) who attempted to extend the work of Roche (1847) on disruption limits from fluid to solid bodies. The former speculated that solid bodies were not likely to tidally disrupt at any distance from a planet and that this was at odds with the formation of Saturn's rings or asteroids breaking up around Earth and Jupiter, whereas the latter had formulated orbital limits for fluid, or strengthless, bodies on orbits around planets. The discovery of Comet Ikeya-Seki in 1965, and its breakup during a close passage by the Sun, prompted further suggestions of strengthless small bodies (Öpik 1966). Similarly, as the orbital and size distributions of the Main Asteroid Belt came into view, collisional evolution via catastrophic disruption was found to likely be energetic enough to have created a large population of shattered leftovers with broken and loosely bound interiors (Chapman 1978). It was in this context that Chapman 1978 first used the term "rubble pile".

Population studies, specifically those that observed brightness variations over time to derive rotation rates (known as "lightcurves"), found a lack of very rapidly rotating bodies (Burns 1975; Pravec et al. 2000,2002). The critical spin limit, which no sizable (larger than ~200m) asteroid exceeded, was ~2.2 hours (see Figure 2). This is similar to the rotation rate at which free particles could leave the surface of a spinning body of density similar to most meteorites. The absence of bodies with rotation rates faster than the observed limits argues that their internal structures do not allow it, but does not mandate it. This useful and highly suggestive dataset, with spin and shape as a combined constraint will be referred to later as a tool to estimate allowable shape and spin configurations relating to granular flow properties of known materials and to constrain their internal structures.

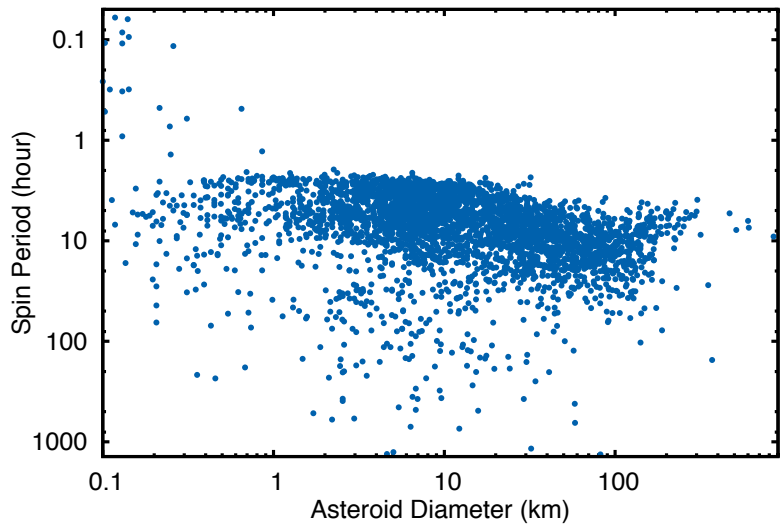

[**Figure 2:** The distribution of observed asteroid spin periods plotted as a function of their diameter (km). This plot relies on the asteroid lightcurves recorded in the Planetary Data System archive (Harris et al. 2016) that have U=3 rating suggestive that they are very reliable. The two outliers, bodies with D>200m and P<2h are 2001 OE84 (Pravec et al. 2002) and 2001 VF2 (Whitely et al. 2002).]

The surfaces of many planets and moons record information about the population of small bodies throughout the Solar System seen through the lens of crater formation. Crater chains in particular helped to drive the idea of fractured or rubble-ized interiors for small bodies that literally fall apart before impact, and helped to lead the way in understanding population-wide properties of small bodies. Crater chains, or "catenae", are linear features of individual impacts that can span tens of kilometers, and can be distinguished from possible local or endogenic sources (Figure 3). The Moon and the Galilean satellites have ideal and well-studied surfaces for these processes due to their proximity to a planet capable of tidally disrupting an asteroid or comet (where Saturn's low density make a violent disruption difficult - Asphaug & Benz 1996). The Moon has a few such crater chains (Melosh & Whitaker 1994, Wichman & Wood 1995) and Callisto and Ganymede have several on Jupiter-facing hemispheres (Schenk et al. 1996). The rate of expected events capable of making the observed crater chains pointed to a large population of strengthless bodies (Bottke & Melosh 1996).

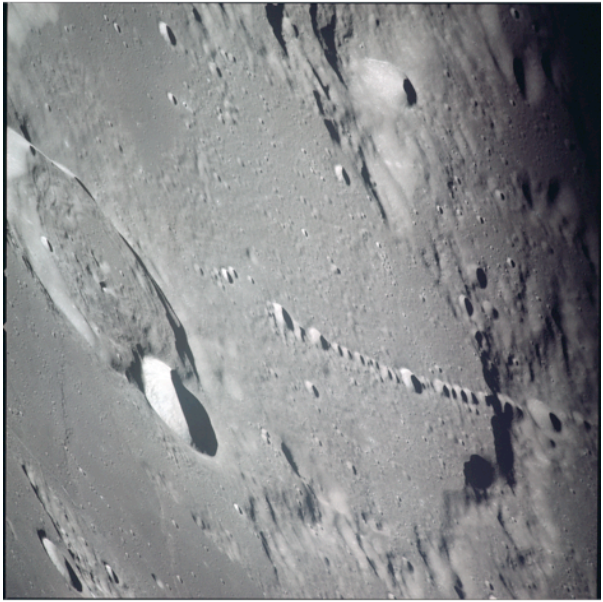

[**Figure 3**: The Catena Davy on the Moon, as captured in an image from Apollo 12. Photo number AS12-51-7485.]

      A key difference between a rubble pile and an intact but broken "shattered aggregate" is the expectation of void space that should be a measureable bulk porosity. This is far from a trivial measurement to make, as knowledge of both mass and volume is required, which is a rare set of data to have for smaller solar system bodies. However, where the measurements exist, the occurrence of asteroids with densities significantly smaller than the grain densities for the most similar meteorites, has regularly pointed to moderate to high porosities for many asteroids, but almost without exception for smaller asteroids.

      In 2004, the world finally saw a rubble pile up close, when JAXA's Hayabusa spacecraft visited the ~500m asteroid Itokawa, surveyed it, and even touched it and returned samples. Images of this asteroid have come to define what it means to be a rubble pile. Next up, to test and exercise these ideas and concepts are the upcoming space missions by both JAXA and NASA to visit, study and return samples from two more asteroids of similar size. Asteroids Ryugu and Bennu will be visited in the coming years, and both are much darker and spectrally distinct from Itokawa, belonging to the taxonomic groups that are suggestive of links to the more primordial and carbon-rich meteorites. All the data in this chapter suggest that both mission targets will be rubble piles and the community hopes to learn as much or more from them as was learned from Itokawa.

# 2.0 Where do rubble piles come from?

This is a two way street – big picture models of solar system evolution provide context important for understanding the character of small asteroids and comets, while properties of asteroids or groups of asteroids inform the big picture models of solar system formation and evolution. In this section solar system context is provided such that planetesimal formation models, asteroid belt collisional evolution and collisional physics conspire to suggest that there should be a population of rubble pile asteroids. However, the body of evidence about the nature of small asteroids has been building since the term "rubble pile" was coined in 1978 and predates much of the recent advances in Solar System evolution modeling.

## 2.1 Planetesimal formation and solar system context

The story of the Solar System's small bodies begins with one of the hardest problems in planetary science - planetesimal formation. Where, how and by what primary processes the first km-scale solid bodies in the Solar System are formed remains a generally open question (see recent review by Johansen et al. 2015). Several lines of evidence suggest that asteroids were "born big" (Morbidelli et al. 2009, Bottke et al. 2005, Johansen et al. 2015, Delbo et al. 2017), with sizes around one hundred kilometers, which is larger than what would today be considered to be the nominal size range for rubble pile asteroids.

The dynamics of the gaseous solar nebula will damp orbits and produce low relative velocities between dust grains and small particles (<cm), which allows growth by sticking. But if particles were able to grow to sizes approaching one meter, aerodynamic drag from the gas disk would be maximized and result in significant loss of material due to rapid inward drift. Furthermore models of grain growth confront other barriers to growth whereby small mm-sized grains may simply bounce or fragment instead of sticking, effectively limiting the maximum size of simple growth at cm-sizes (Zsom et al. 2010, Wurm et al. 2005, Blum and Wurm 2008, Johansen et al. 2015). Therefore models rely on various gas-disk processes related to turbulence and instabilities to quickly collapse swarms of the mm- to cm-sized particles that dominate the texture of most primitive meteorite samples and quickly form the much larger planetesimals (Cuzzi 2008, Johansen et al. 2007; Johansen et al. 2015).

The concept of a preferential initial minimum planetesimal size comes from both formation models and the analysis of the current asteroid size frequency distribution. An initial, characteristic, planetesimal size is fossilized in the current asteroid belt size distribution as a "bump" or a "knee" at D~100km (see Fig. 4; Bottke et al. 2005,2015; Morbidelli et al. 2009). This signature remains while some of the original population have suffered collisions, broken into smaller fragments and helped to populate the distribution at smaller sizes. This collisional evolution could drive an original belt entirely devoid of small objects (smaller than the characteristic D~100km) to its current state on billion-year timescales (Bottke et al. 2005). These studies suggest two important implications for everything smaller than this characteristic initial size:

1. They are not likely first generation, or original, members of the asteroid belt.
2. They likely formed as part of the collisional evolution/disruption of a larger body.

Constraints on both points can be solidified by understanding the collisional environment in the current asteroid belt. Collisional probabilities found in the current asteroid belt suggest that large collisions happen somewhat regularly (on Solar System timescales), and this is seemingly confirmed by the known suite of collisionally-formed asteroid families and the highly brecciated properties of many different types of meteorites (Bottke et al. 2005, 2015, Nesvorny et al 2015, DeMeo et al. 2015, Housen et al. 1982). Calculating expected timescales between catastrophic collisions for asteroids of a given size requires both the collisional probabilities and the physics that govern their shattering and dispersal.

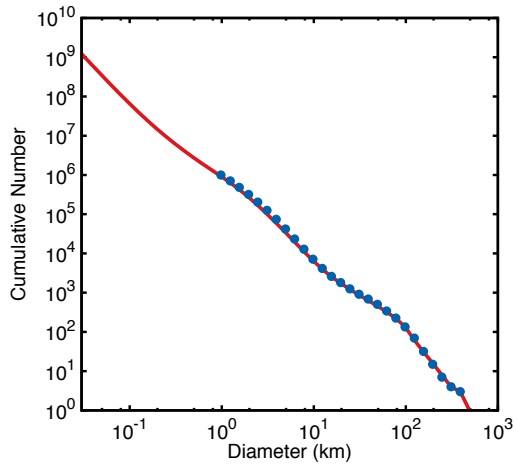

[**Figure 4**: The size frequency distribution of the Main Asteroid Belt. Of note is the structure of the curve around a few hundred meters and one hundred kilometers that is typically interpreted as relating to the size of the transition to strength-dominated bodies, and the size of the original population. Data from Bottke et al. 2005, 2015.]

## 2.2 Collisional evolution - reaccumulating to form rubble piles

Shattering and dispersal are essential steps in the larger concept of "collisional evolution" of asteroids, but so is the reaccumulation of their fragments. Shattering is something only applicable to a body with some bulk material strength where the required energy to shatter is typically measured with a value called the critical specific shattering energy, $Q^*_s$. This can be probed in terrestrial laboratories at cm-sizes and then at larger scales in hydrodynamic simulations (Nakamura et al. 1991, Benz & Asphaug 1999). $Q^*_s$ specifies the projectile kinetic energy per mass required to shatter the target to the point where the largest fragment contains 50% of the system mass (Durda et al. 1998). The energy per mass required to shatter a body decreases as the targets get larger due to the internal flaws increasing with size (Housen & Holsapple 1990; Holsapple 1994), and this trend holds up to the point that gravity becomes more important than strength.

Simply shattering bodies will not evolve the size distribution of the entire main asteroid belt - they must be shattered and dispersed to change the population statistics. The critical specific dispersing energy is $Q^*_d$, which is the energy required to leave a largest remaining remnant that is only 50% of the original target mass. This value increases with size as gravity grows more important, but at small sizes it is a bodies' strength that determines collision outcomes. The crossover point of strength and gravity regimes for unfractured basalt targets is estimated to be around D ~300m (Asphaug et al. 2002) and is reflected in another bump/wave in the size frequency distribution of the asteroid belt that is expressed at sizes around 1km (Bottke et al. 2005). This size is subject to many uncertainties regarding the possibly pre-fractured state of the target before its impact (Benavidez et al. 2012) and the averaged effect of different and potentially oblique impacts (Movshovitz et al. 2016) or pre-impact rotation (Ballouz et al. 2014). Furthermore, as is discussed below as part of the conclusions about internal structure deduced from shapes and spin, this size range is where bodies are found to be spinning more rapidly than what would be allowed for a body bound only by gravity – generally suggestive of some strictly monolithic strength-dominated bodies at and below this size.

When bodies are disrupted their debris not only disperses as individual strength-dominated monolithic fragments, but the dispersed fragments themselves are expected to reaccumulate into a series of new asteroids (Michel et al. 2001). This process has been constrained by the size distribution of reaccumulated fragments compared to observed collisional families (Michel et al. 2001, 2002 etc. Durda et al. 2004, Benavidez et al. 2015), and the formation of satellites (Durda et al. 2007). This points directly to the origin of rubble pile asteroids - by the reaccumulation of fragments following a catastrophic asteroid collision.

This reaccumulation process is difficult to model - as a huge number of fragments might be expected to reaccumulate following a giant impact. The actual impact and shockwave propagation through a target requires hydrodynamics simulations to track the shock propogation through a solid target, but once the target is shattered the dispersal and reaccumulation are governed by their gravitational attraction and relatively low-speed impacts. Therefore most efforts include a handoff from hydrodynamics to *N*-body gravity particle codes (Michel et al. 2001,2002,2003,2004, Durda et al. 2004, 2007, Benavidez et al. 2015). The latter steps are numerically expensive and one of the many simplifications taken to keep simulations reasonably sized is to perfectly merge the particles reaccumulating into each individual fragment. While it is clear that all of the particles will form a new reaccumulated asteroid matching the size frequency distribution, no shape or spin information is retained to study the individual bodies in the newly built family of asteroids (see Figure 5; Michel and Richardson 2013, Walsh et al. 2017).

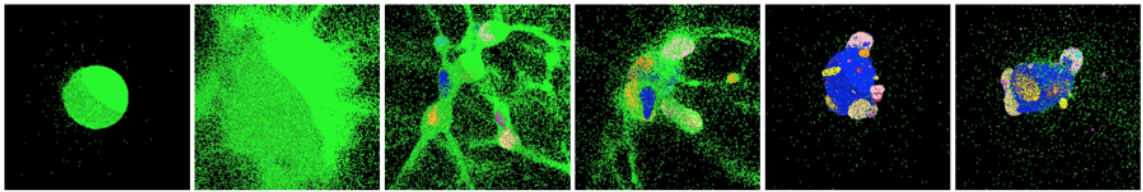

[**Figure 5:** The reaccumulation of fragments following an asteroid catastrophic collision, with a new numerical technique that keeps all particles in the system. Credit to Michel and Richardson 2013]

Collisional lifetimes, the expected time between catastrophic impacts for an asteroid of a given size, can be calculated with $Q^*_s$ and $Q^*_d$, and the orbit and size distribution of the asteroid belt (Bottke et al. 2005, 2015). The expected timescale between disruption events for a body increases with size in the gravity regime due to both increasing $Q^*_d$, due to increasing gravity and the decreasing number of available impactors massive enough to disrupt them. A ~10 km body has a collisional lifetime similar to the age of the Solar System, and thus anything below this size should not have survived in its initial state (Bottke et al. 2005). This size sets the upper limit for what is typically expected to be a rubble pile as everything smaller than this is likely a reaccumulated remnant, and larger than this size *could* have survived roughly intact.

Combined, expectations from formation and evolution models point to a size range between ~200m-10km where a reaccumulated rubble pile structure should predominate. The lower end of the size range for rubble piles is deduced from the $Q^*_d$ curve and the transition from gravity dominated regimes to strength dominated regimes and is supported by the distribution of known spin rates to be ~200m. The large end of this range is deduced from collisional lifetimes, modeling suggesting an initial characteristic size D~100km, and is supported by the bump in the size distribution at these larger sizes. Spacecraft visits provide some insight as D~34km asteroid Eros is considered more likely a shattered, but intact, aggregate rather than a re-accumulated rubble pile based on its density and surface geology (Cheng et al. 2002), while ~350m Itokawa is clearly a rubble pile.

# 3.0 Internal Structure

By understanding the internal structure of rubble pile asteroids we aim to build constraints on their place in Solar System evolution, to understand how they have behaved and evolved over time, and to learn how to model them and interact with them now and in the future. Nearly all data on hand are astronomical - population data about shapes and spins and thermal and radar data about the surfaces. For a handful of objects there is actual resolved images of surfaces from flybys and rendezvous. These data constrain the modeling efforts that range from semi-analytic models of surface material flow to N-body models of gravitational bound particles with a variety of surface forces. Combined they build our picture of a rubble pile's internal structure.

The critical findings from the data and models suggest that while high porosity and a lack of tensile strength may be a defining property of a rubble pile, their shear strength and a resistance to re-shaping is extremely important for defining the state and behavior of a rubble pile. The importance of shear strength is clarified in a simple example: while neither water nor sand on the beach has any tensile strength, the shear strength of sand is why we can walk on it and not water. Likewise, rubble piles show minimal resistance to disruption by a primarily tensile force such as tidal disruption, but if slowly spun to rapid rotation rates they may suffer only minimal re-shaping and can maintain spheroidal shapes. The attempt to characterize these properties, and the desire to model it, have driven much of the research in this field.

### 3.1 Density, porosity and satellites

A key trait for a rubble pile asteroid is the presence of void space, or macro-porosity, that is inferred by way of low density compared to analog meteorites (where any significant micro-porosity would seemingly be found in meteorite analogs). The first glimpse inside an asteroid is attained by way of measuring its density, but the bulk density calculation for an asteroid requires its mass and volume. The mass itself can only be measured in a few ways with the most widely used being orbital deflections, where asteroids can alter the orbits of other asteroids (see Carry et al. 2012). This can attain accuracies of a few percent for the largest asteroids but rapidly decreases in accuracy for smaller sizes. Spacecraft that have flybys with an asteroid can measure mass by way of the detected trajectory deflection or directly from the orbit, although this has been done for only a small handful of objects (Yeomans et al. 1997, 2000; Fujiwara et al. 2006, Patzold et al. 2011). For objects with both a shape measurement and a decades long measurement of its orbital drift rate by the thermal Yarkovsky effect, which is dependent on mass, a mass and density estimate can be made, as it was done for near-Earth asteroid Bennu (Chesley et al. 2014).

Meanwhile, the properties of a satellite's orbit can be used to attain the system mass and has the advantage of being equally effective for large and small asteroids, and with the population of ~250 known asteroids with satellites being spread throughout a wide range of dynamical groupings and asteroid sizes, this is the key technique for measuring the density of rubble pile asteroids. Depending on the discovery technique(s) the information come with some uncertainties, but in some cases they have rivaled anything returned from a space mission. The first discoveries of asteroid satellites around small asteroids were made with lightcurve observations - brightness fluctuation over time - where in the case of eclipses and occultations the existence of a secondary can be deduced (Pravec & Hahn 1997, Mottola & Lahulla 2000). Discoveries of satellites made with this technique can relay information on the primary mass and density with various assumptions about the relative density between the two components and the system geometry, where Pravec & Hahn (1997) first reported a density of $1.7 \pm 0.4$ grams cm$^{-3}$ for small near-Earth asteroid 1994 AW1.

Meanwhile, radar detection and characterization of binary systems has provided a wealth of data, directly providing constraints on orbits and absolute sizes of each component with one set of observations. Radar also provided valuable confirmation of the lightcurve discovery techniques after observing some of the same targets (Margot et al. 2002). Combined, the two techniques found that about ~16% of all near-Earth asteroids have satellites (Margot et al. 2002), and have produced densities for 13 NEAs with diameter below 10km, estimating an average porosity of $35^{+38}_{-35}$% (Carry et al. 2012). The object with the smallest uncertainties is 1999 KW4 with porosity of $45\% \pm 16$, which is comparable to the value derived by the spacecraft visit to asteroid Itokawa, $42\% \pm 11$ (Ostro et al. 2006; Fujiwara et al. 2006).

Overall, these data point to significant void space and high porosity on average among NEAs in this small size range. For context, hexagonal closest packing of spheres is ~74% efficient, leaving 26% void space, while irregularly, or jammed, packing of spheres is ~63% efficient with 37% void space or porosity (Song et al. 2008). The average value found here is similar to the irregular packing, while a handful of NEAs have values that actually exceed 50%, pointing to a potentially complex interior structure. It is notable that the large NEA Eros (D~34km), also visited by spacecraft, has an estimated porosity of $19\% \pm 1$, which is part of the reasoning that drives its inclusion in the grouping of shattered aggregates rather than rubble piles (Carry et al. 2012; Cheng et al. 2002).

## 3.2 Rotation rates and shapes constrain internal structure

Seeking brightness fluctuation over time, lightcurve observations and data reduction have become increasingly automated over time resulting in a large database of objects (1757 with quality code U=3 with 200m < Diameter < 10km in the PDS archive v16; Harris et al. 2016). A single apparition of an asteroid can provide very precise data on an asteroid's rotation rate from the periodicity of the lightcurve but only weak constraints on its elongation from its amplitude due to the unknown geometry of the rotation axis relative to the observer. As the number of apparitions increase, the ambiguities about spin axis direction can be resolved and tighter constraints on shape and elongation can be derived and eventually shapes can be constrained using lightcurve inversion techniques (Durech et al. 2010, 2015).

Given the relative ease of establishing a rotation period for an object, first-order analyses of the entire population are typically made assuming the end-member case of a spherical body. This drives the common comparison seen in Figure 2, where the singular variable of spin rate is found to not exceed what would allow a surface particle to escape from the equator of a sphere. Long the anchor for presumptions of a strengthless interior this compelling relationship does not demand any particular interior structure but is a compelling argument for a lack of strength. There are only two well-characterized outliers, 2001 OE84 (Pravec et al. 2002) and 2001 VF2 (Whitely et al. 2002), although more have been suggested in a range of different surveys (Chang et al. 2015).

Lightcurves also provide information about axis ratios, where the lightcurve amplitude during one apparition provides constraints on the equatorial $b/a$ amplitudes, where $a,b,c$, are the long, intermediate and short principal axes respectively, and principal axis rotation around the $c$ axis is assumed. Multiple apparitions can constrain the spin pole direction and eventually all three axis lengths. After converting lightcurve amplitude from observed magnitude to an estimated axis ratio ($\Delta m \sim 2.5\log(a/b)$ ; note that other effects such as albedo variations can also affect brightness fluctuations so that this does not provide a perfect indicator of shape), there is an added dimension to the critical spin period comparison, where expected maximum spin rate will decrease with increasing elongation ($P_{crit}/3.3\text{hrs} = \sqrt{((1+ \Delta m)/\rho)}$). Figure 6 shows that with the critical spin period is still respected for the entire population of asteroids, for asteroids with 200m < D < 10km.

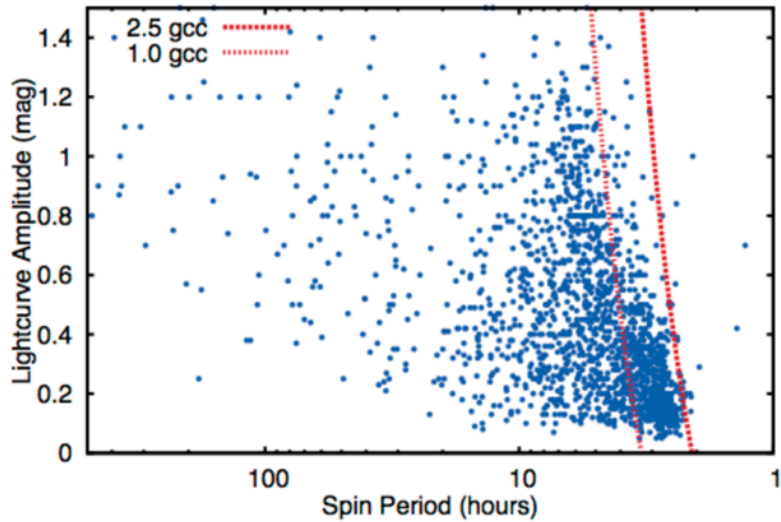

[**Figure 6:** Lightcurve derived amplitudes in magnitude plotted as a function of their spin periods (hours). The lines represent critical spin periods considering the lightcurve amplitude. Note that the maximum lightcurve amplitude is taken from the PDS archive, and geometry concerns only set this as a lower limit, which could result in moving some data points up (Harris et al. 2016).]

The simple fact that there are fast spinning spheroidal objects in this population demands that they are not behaving like fluids. A fluid, following the equilibria pathways described by Jacobi and MacLaurin, would immediately begin re-shaping after attaining non-zero angular momentum. Thus, the entire set of bodies in Figure 6, if they were acting as fluids should actually fall on a line, rather than fill a large space, and none should be in the lower right corner where rapidly spinning spheres are found. This all points the population having some amount of shear strength to be able to maintain non-fluid shapes while rotating (Farinella 1981, Holsapple 2001, Tanga et al. 2009, Richardson et al. 2005). This behavior relates quite naturally to a granular material, like gravel, where inter-particle friction and particle interlocking provides a shear strength that prevents reconfiguration of the materials. In simplest terms this is expressed in a materials "angle of friction" or "angle of repose", where it can withstand slopes up to a critical angle beyond which it flows.

Treating a rubble pile as a granular material opens up numerous modeling and characterization avenues. One such technique modeled rubble piles as cohesionless elastic-plastic solids, characterising the allowable shape and spin combinations, which can be outlined by an envelope in terms of axis ratios and rotation rate (Holsapple 2001, 2004). For the Mohr-Coloumb yield model the size of the allowable envelopes depends on the materials angle of friction, which is the primary input parameter (Holsapple 2001,2004). With this analysis the known data on asteroid spin states can be compared against allowable envelopes for a range of angle of friction (note that many authors have also deployed a Drucker-Prager model; Holsapple & Michel 2006, Sharma et al. 2009, Rozitis et al. 2014). But, comparing the observations to the predicted envelopes is challenging due to widely varying densities, which are typically unknown and required for normalizing the observed rotation rates. Furthermore, a handful of outliers have been identified that absolutely must be bound by more than just gravity. Largely, this analysis finds that nearly all small asteroids could have their shape and spin configurations explained if their constituent material has an angle of friction similar to most known terrestrial materials, around 40 deg.

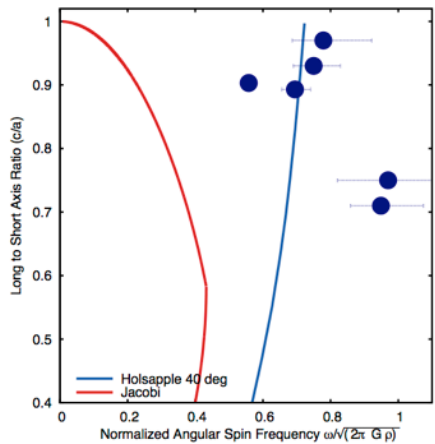

**[Figure 7:** Asteroids (from left to right), Bennu, 1999 KW4, 2001 SN263, 1994 CC, DP107, 1950 DA (Lauretta et al. 2014, Ostro et al. 2006, Rozitis et al. 2015, Naidu et al. 2015, Becker et al. 2015). Note that 1950 DA was shown to require some cohesive strength to maintain its shape and spin configuration (Rozitis et al. 2014). All of the objects, except Bennu, have satellites, and 2001 SN263, 1994 CC are both triple systems.**]**

Individual bodies can be analyzed this way and compared directly against fluid equilibria and cohesionless spin limits (see Figure 7). A handful of near-Earth asteroids have very well-characterized shape models, densities and rotation rates (these are typically the primary of a binary system). They are typically found to have top-shapes or spheroidal shapes with equatorial ridges, which is discussed in depth below as they are suspected to have originated during an episode of spin rate increase. The comparison between their spin/shape configuration and these fiducials finds three of these six objects close to the 40 deg friction line of Holsapple (2001), two are well-beyond this limit, and Bennu is safely inside the 40 deg friction envelope. Of the two that are beyond the 40 deg friction limit, one is 1950 DA which has been specifically modeled to have some cohesive strength (~62 Pa; see Rozitis et al. 2014). The minimal cohesion necessary to explain 1950 DA being so far beyond the allowed friction limits shows the difficulty with this type of analysis - even miniscule amounts of cohesion by terrestrial standards can quickly allow spin rates well beyond the limits allowed by simply friction and gravity.

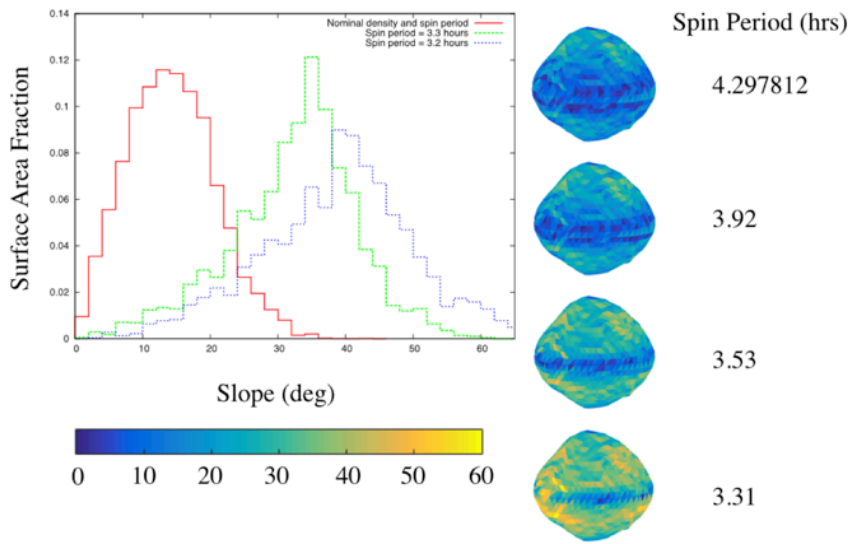

[**Figure 8**: The slope distribution across the surface of the asteroid Bennu at a range of spin states, where histograms show the fraction of the surface facets at a given slope for different spin rates of the asteroid. The slope distribution at the different spin rates is visualized in the images to the right following the colorscale on the bottom. Plot taken from Scheeres et al. (2016).]

On the other end of this group is the asteroid Bennu, which has the same top-shape found for all other objects in this group but is not near an extreme spin rate (see Figure 8). Here, if assumptions about the material properties are made - that they are similar to the other objects - then the distribution of surface slopes across its surface can be tracked using its spin rate as a free variable. In order to match slope distributions found on other rapid rotators, with peaks between 30-40 deg., a spin rate of 3.3 hr is needed, which is significantly faster than the current ~4.3hr rotation rate. As will be discussed below, there are multiple mechanisms that can change spin states or shapes of near-Earth asteroids of this size, so this analysis of Bennu suggests a past history of more rapid spin and re-shaping.

### 3.3 Odd shapes and how to reshape a rubble pile

While the shape and spin configurations of rubble piles provides insight on their material properties they can also point to evolutionary mechanisms that have helped shape the population. These small bodies are subject to significant forcing due to interactions with planets (NEAs) and solar radiation - providing valuable laboratories to understand them better. One of the most distinct features observed among small near-Earth asteroids are the spheroidal bodies with top-shapes or equatorial ridges. This particular shape is not trivial to determine - a lightcurve observation would witness only minimal brightness fluctuation owing to the nearly circular equator at all apparitions. Here the capabilities of delay-doppler imaging via radar has been able to interpret these shapes (Ostro et al. 2006). The width of the doppler returns track surface area at different rotation rates about a spin axis. The delay imaging follows the timing of returns and provides topography relative to the sub-earth point on the surface. Thus in the delay imaging the topography of the ridge is different than what would be returned from a perfect sphere. This shape is commonly found among the population of rapidly spinning asteroids that have satellites, suggesting that this is an end-state of a spin-up and binary formation process (Walsh et al. 2008, Harris et al. 2009, Scheeres et al. 2016), but is also found around at least one non-binary with moderate spin - Bennu the target of NASA's OSIRIS-REx mission (Lauretta et al. 2015,2017).

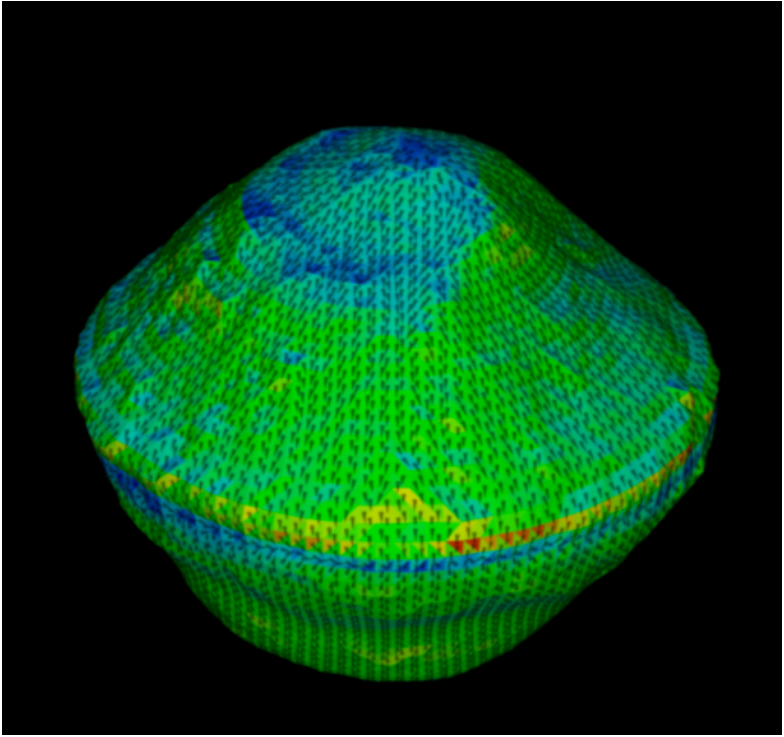

[**Figure 9:** Near-Earth asteroid 1999 KW4, is found to have average surface slopes of 28 deg and maximums of 70 deg. It has a rapid rotation rate very near critical, and a satellite in a close and nearly circular orbit (Ostro et al. 2006, Scheeres et al. 2006). The colorscale shows slopes (blue is low and red is high, with the mid-latitude greens being ~30deg) and the arrows show the steepest descent.]

The other end of the spectrum are highly elongated, bifurcated and contact binary asteroids (Benner et al. 2015). Benner et al. 2006 estimated that ~10% of NEAs > 200m are contact binaries, following a strict definition for contact binary, where two lobes and bi-model mass distribution is required. Note that this definition would exclude objects such as Itokawa and Toutatis that have a shape suggestive of multiple components, but at a higher mass ratio between components (Fujiwara et al. 2006, Zhu et al. 2014). Using a relaxed definition that allows mass ratios between components as low as 4:1, which includes Itokawa and Toutatis, the fraction of the NEA population grows to ~14% (Taylor et al. 2012).

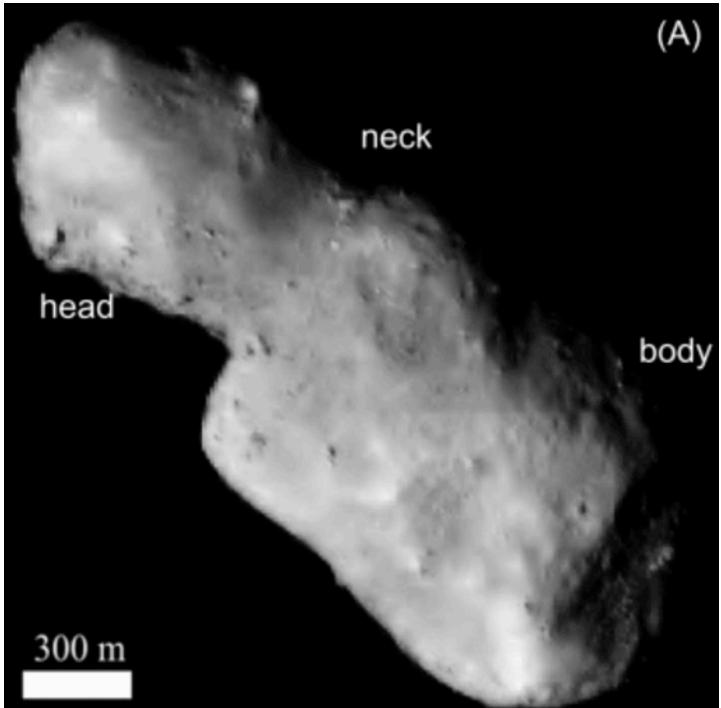

[**Figure 10:** The asteroid Toutatis is a km-scale NEA that shows a bifurcated shape. It was visited by the Chang'E-2 spacecraft (Zhu et al. 2014).]

The first effective dynamic models of the full-body behavior of rubble piles were motivated by the disruption of comet Shoemaker-Levy 9 (SL9) in 1992. Spectacular observations showed tens of fragments of the comet strung out along its orbit (Weaver et al. 1994), where the orbit traced backwards to a closest passage by Jupiter at only 1.31 Jovian radii (Yeomans and Chodas 1994). Such a close passage was likely within the Roche Limit, which is a classical disruption limit used widely to interpret close tidal interactions. However, the breakup was interpreted to have been quite violent given the large number of coherent and similar-sized fragments that were observed with no particular dominant member (Weaver et al. 1994). This event showed the inadequacies of simple disruption limit calculations such as the Roche Limit, which itself is built on assumptions of fluid equilibria, and instead demanded calculations and models more conscious of the interior dynamics of a body constructed of irregular shaped, and gravitationally bound, constituent pieces (Asphaug & Benz 1994,1996; Solem & Hills 1996, Richardson et al. 1998).

The first numerical models born from these studies were billiard ball, *N*-body, type models, those capable of modeling a gravitational aggregate, a body held together entirely by its self-gravity, but without some of the complicating dynamics introduced by surface penetrations and interactions of the constituent particles (Asphaug and Benz 1994,1996; Richardson et al. 1998,2000). Such a model was well-suited to capture the dynamics of tidal forces pulling an asteroid apart in a relatively impulsive event.

### 3.3.1 Tidal Disruption

The disruption limits and behavior of a simple rubble pile, in the absence of any tensile strength, only scale with density and not absolute size (Solem 1994), so models performed for SL9 can be applied to near-Earth Asteroids encountering the Earth and Venus or Trojan asteroids flying past Saturn and the outer planets (Asphaug & Benz 1996). The starting point for these works is the classical Roche Limit, which derived the orbital distance at which a body would disrupt around a planet (Roche 1847). This limit is a valuable fiducial but was derived assuming a fluid body with a specific axis ratio and orientation relative to the orbit.  Analytical modifications were made to account for realistic failure of brittle materials (Dobrovolskis 1990) and for a viscous body on a parabolic encounter, which substantially lowers the needed close approach distance for breakup (Sridhar and Tremaine 1992). These limits are simply functions of the relative density between the progenitor and the perturber or planet, but fundamentally describe at what distance something may disrupt, but not address how the magnitude or violence of the breakup nor the morphology of the resulting fragments nor can they easily account for even simplistic spin states or irregular shapes.

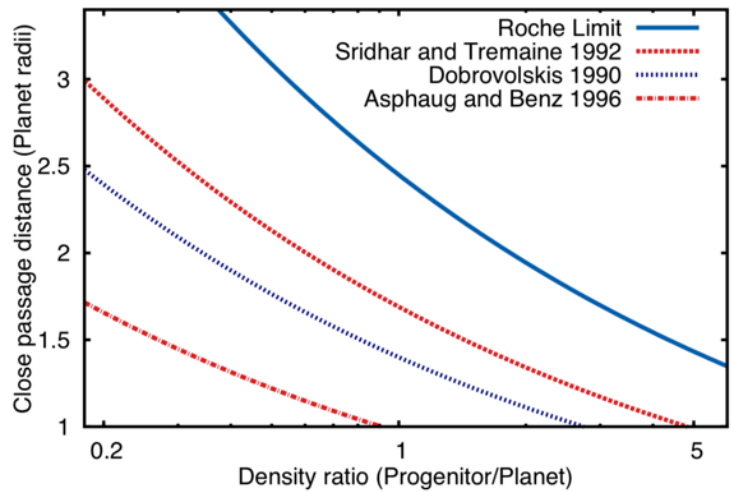

[**Figure 11:** Limits for tidal disruption plotted as the close approach distance to a planet in planetary radii as a function of the density ratio between the small body and the planet. The four calculation shown are the classical Roche Limit, the flyby limit from Sridhar and Tremaine (1992), the limit derived by Dobrovolski (1990), and the simulation outcomes from Asphaug and Benz (1996) for a SL9-type disruption.]

SL9 didn't just disrupt, it *really* disrupted. The imaging campaigns provided solid constraints for further modeling based on the fragment chain, and this was the launching point for the *N*-body approach to modeling rubble piles. The morphologic outcomes played a large role in understanding what the outcome should look like as a function of relative densities between progenitor and planet, which was constrained in a series of works to be quite low (~0.6 grams cm$^{-3}$; Asphaug & Benz 1994,1996). Furthermore, these models allowed for a more detailed view of disruption, where the magnitude, or violence, of a disruption could be established and limits set for each. Richardson et al. (1998) built a basic descriptive taxonomy for levels of disruption based on mass loss:
- S-class, or catastrophic, disruption named after Shoemaker-Levy 9, where more than 50% of the progentior mass is lost,
- B-class disruption, where the largest remnant has between 50-90% of the total mass,
- M-class disruption that is mild where less than 10% of mass is lost.

The SL9-type, or S-class, disruptions, require a density difference between the progenitor and the planet (see Figure 11). For the scaling found, an SL9 disruption would not be possible at Saturn for a comet with similarly low density as Saturn (Asphaug & Benz 1996), whereas the terrestrial planets provide a large density difference than most small bodies and are good candidates to disrupt rubble piles.

The new modeling capabilities also permitted investigations into a wide range of spin states and shapes of the disrupting body (Asphaug & Benz 1996, Richardson et al. 1998). Some important properties are degenerate during a tidal disruption - more rapid rotation in the prograde direction led to more violent breakups while retrograde rotation frustrated breakup. Thus rotation and close approach distance had some degeneracy, and the direction of the spin axis and the long-axis during close approach could both further complicate the outcomes (Richardson et al. 1998). Furthermore these efforts also clarified that even small amounts of tensile strength, provided by way of some sort of structural cohesive bonding, could seriously alter these outcomes and frustrate disruption (Asphaug & Benz 1996, Holsapple & Michel 2008).

Close tidal encounters have the capability to distort shapes, even in the case of no mass loss. The particular case of near-Earth asteroid (1620) Geographos stands out due to its highly elongated shape with irregular cusped ends and the close matches found in a simulation (see Figure 12; Bottke et al. 1999). How common this specific shape is throughout the population of bodies that can possibly encounter a terrestrial planet is not easily known, but the frequency of contact-binaries (~10%) and bifurcated shapes (~14%) have been quantified and are roughly as common as binary systems (Benner et al. 2006, Taylor et al. 2012). There are distinct modeling outcomes that produce shapes similar to Geographos, where a close approach of an ellipsoid with prograde rotation relative to its encounter experiences significant distortion and minimal spin rate increase, and ends up as a perfect doppelganger for Geographos (Bottke et al. 1999).

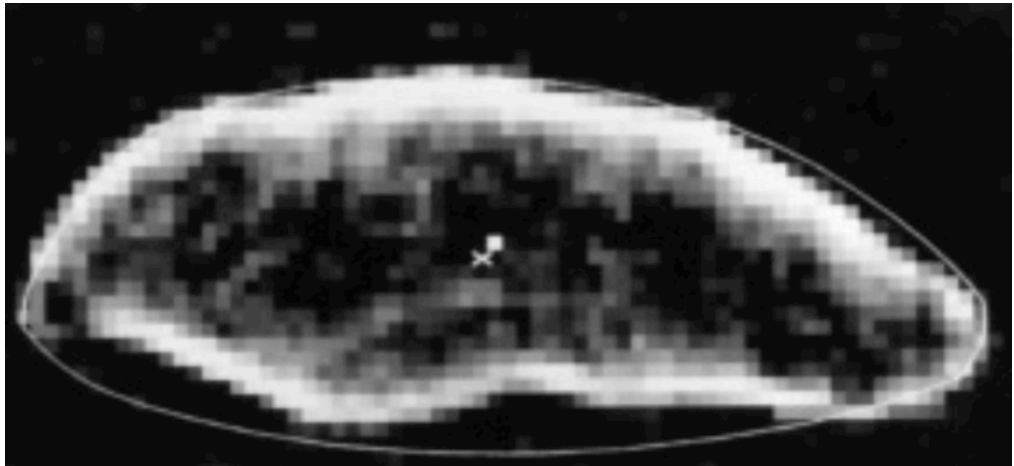

[**Figure 12**: Near-Earth asteroid Geographos was found to have a highly irregular shape, which was matched in N-body simulations of a tidal disruption (Bottke et al. 1999).]

Including more realistic surface interactions between the constituent particles, by way of a soft-sphere discrete element model (SSDEM; see Sanchez & Scheeres 2011,2012; Schwartz et al. 2012; Zhang et al. 2017), actually generate more highly elongated shapes than found in the first generation of tidal disruption models (Walsh et al. 2017). In some cases a few pieces in the fragment train are bound due to the slight increase in frictional forces during disruption (a contrast to some previous models with minimal or no friction) and result in very low velocity collisions between fragments leading to some bifurcated shapes that resemble asteroids such as Itokawa, Toutatis etc., and could possibly account for some of the large population of contact binaries.

### 3.3.2 Spin-up by the YORP effect

The population of near-Earth asteroids with satellites makes up ~16% of the entire near-Earth asteroid population (Margot et al. 2002, Richardson & Walsh 2006, Margot et al. 2015, Walsh & Jacobson 2015). The primaries are typically very fast rotators (periods < 4hr), and the satellites are small and close (10-40% the size of the primary and 2-4 primary radii away). While tidal disruption was shown to produce binary systems (Richardson et al. 1998, Walsh & Richardson 2006), they were not great matches for these properties and lightcurve surveys found that a population similar to the near-Earth asteroids existed in the Main Asteroid Belt - far from any planet that was capable of tidally disrupting any asteroid (Walsh & Richardson 2008, Pravec et al. 2006). Another mechanism was needed to explain this population and Bottke et al. (2002) proposed that a thermal affect similar to the Yarkovsky effect that changes orbits, known as the Yarkovsky-O'Keefe-Radzievskii-Paddack effect, or YORP-effect, might be capable of spinning asteroids to disruption (Rubincam 2000, Bottke et al. 2002).

[**SIDEBAR**: The YORP-effect relies on the asymmetries of an asteroid's shape to produce a torque around its spin axis when it reflects or re-emits solar radiation (Rubincam 2000, Bottke et al. 2006). The action of the YORP effect is directly observed in a handful of objects that are observed to be actively spinning up (Taylor et al. 2007, Kaasalainen et al. 2007) and long-term evolution is found in populations of asteroids that are driven into spin-orbit resonances in the Main Belt ("Slivan States" - Slivan et al. 2002; Vokrouhlický et al. 2003) and in the way that asteroid families drift apart over time (Vokrouhlický et al. 2006, Nesvorný et al. 2015). The YORP-effect can also alter obliquities, effectively pushing asteroids towards 0 or 180deg obliquities (Vokrouhlicky & Čapek 2002; Čapek and Vokrouhlický 2004; Vokrouhlický et al. 2004). Whether or not a specific asteroid is likely to be spinning up or spinning down is not-trivial to calculate, even for a well-studied body with a known shape, and some studies suggest that even small scale surface artifacts like boulders or very small craters can change how a body responds (Statler 2009). ]

It's hard to overstate how small a given YORP torque would be on a 1km asteroid. The time for a body of this size to halve or double its spin rate (YORP can spin down objects too) by this effect is on order ~1 million years (Bottke et al. 2006), which, while slow, leaves plenty of time during an average near-Earth asteroid lifetime (~10 Myr) or the collisional lifetime of a 1km main belt asteroid (~500 Myr) for significant spin-state alteration. As implied by the spin and shape configurations of the entire population of asteroids there is an expectation of some shear strength acting in the form of a simple angle of friction - something to prevent these bodies from reverting back to fluid equilibria (Holsapple 2001; Richardson et al. 2005; Walsh et al. 2012). Spinning bodies up provides a way to slowly change angular momentum while maintaining the axis ratios of the bulk figure.

Models of granular flow can generate first-order estimates of bulk shape changes under increased rotation where the effective slope angle on a patch of the surface can dramatically change as the spin rate increases (as the spin approaches the critical 2.2hr limit the effective gravity at the equator goes to zero). As slopes pass the critical angles of repose failure of slopes is assumed and material is allowed to "flow" down to areas with lower potentials (Guibout and Scheeres, 2003; Walsh et al. 2008; Harris et al. 2009; Scheeres 2015). The bulk shape of 1999 KW4 is closely matched when the body has an angle of friction of 37°, which meets the expectations from the spin and shape distributions for all small asteroids (Harris et al. 2009). Failure is dominant in the mid-latitudes on spheriodal bodies, with material moving towards the equator, creating the observed equatorial bulge (Walsh et al. 2008; Harris et al., 2009; Scheeres 2015).

Similar *N*-body tools used for tidal disruption models were deployed to model how rubble piles would respond to the gradual torques provided by YORP spinup (Walsh et al. 2008,2012; Sanchez and Scheeres 2012,2016, Zhang et al. 2017). Models with simplistic or no surface forces mimic shear strength by hexagonal closest packing of similar sized spheres, or bi-modal distributions of spheres, finding that the spin-up process produces material flow towards the equator of spherical bodies that are qualitatively similar to the observed equatorial ridges observed (Walsh et al. 2008,2012). With better surface forces by way of the soft-sphere discrete element models (SSDEM) that allow modeling frictional material with a high (~37 deg) angle of friction, the results become less straightforward as some failure modes appear internally in the evolving body, rather than just on the surface (Sánchez & Scheeres 2012,2016, Zhang et al. 2017). While some of these outcomes can result in equatorial bulging, they often end up far more extreme than the sometimes subtle bulge seen in so many radar images. One possible solution is that a denser, or stronger, core of material is resisting internal deformation (Walsh et al. 2008, 2012; Hirabayashi et al. 2015; Sánchez and Scheeres 2016). In the first models the effective strength of the "core" was provided by a simple density increase for organized and closely packed spheres surrounded by unorganized and unpacked spheres (Walsh et al. 2008, 2012; Hirabayashi et al. 2015). However, the presence of a strong, or bound, core could come about by fine grains that have settled toward the center of a rubble pile and provide a cohesive binding (Sánchez & Scheeres 2016) or simply by the reaccumulation process preferring to put the largest and most irregular blocks in the interior. Either route to providing a strong interior essentially pushes the failure towards surface shedding modes rather than internal failure.

The connection between the YORP spin-up process, its ability to generate shapes qualitatively similar to the ubiquitous top-shape and the observations that nearly all top shapes have satellites combine to tell a compelling story of rubble pile evolution. One of the oddities in the entire story of YORP spin-up and re-shaping is the asteroid Bennu. It has an equatorial ridge that is found among so many fast-spinning NEAs with satellites, yet its rotation rate is moderate and it has no known satellite. When it is visited and surveyed it will hopefully reveal geologic clues to point towards the origin of this notable shape.

### 3.4 Tidal Dissipation

There are many known rubble piles with satellites where the tidal interaction between the primary and the satellite could provide a means to probe their internal properties. The population of known satellites continues to grow, with more extreme systems continuing to be found (the number of triple systems increased in 2017), while other systems have been studied for decades long timescales (Scheirich et al. 2015). Nearly all known binaries among small, and likely rubble pile, asteroids have strikingly similar properties - with rapidly rotating primaries, close and small secondaries with orbital periods much longer than the rotation rate of the primary (see Margot et al. 2015, Walsh & Jacobson 2015). The gravity of the secondary should raise a tidal bulge on the primary, but since the primary rotation is more rapid than the secondaries orbital period, this bulge is expected to lead the satellite. This bulge can then torque the satellite and evolve its orbit, while slowing the rotation of the primary. The timescale for these processes depend on the internal structure of both bodies and how they dissipate this tidal energy.

The application of any classical tidal evolution calculation could estimate the asteroid rigidity and dissipation factors (in similar terms as used for rocky planets) if the system timescale for tidal evolution is known (Margot et al. 2002). Near-Earth asteroids have relatively short lifetimes compared to Solar System history or typical Main Belt collisional timescales, averaging ~10 Myr. This average lifetime provides a rough timescale for evolution to their current orbital distance, leading to values of rigidity less than rocky planets (Margot et al. 2002). Beyond just the uncertainties in the evolution timescales critical to this calculation, the unique structure of rubble pile interiors could provide a dramatically different dissipation than typically assumed with tidal evolution models. Reformulated based on the consideration of energy dissipation at the few high-intensity points of contact within the body, the effective rigidity of the body decreases and tidal evolution can theoretically proceed much faster than for solid bodies (Goldreich & Sari 2009).

However, confusing any estimation or calculation of a tidal timescale, is the continuing spin state evolution of the primary body caused by the YORP-effect, and its related effect that can evolve the orbit of the secondary, referred to as Binary-YORP, or BYORP (Ćuk & Burns 2005; McMahon & Scheeres 2010). In the same way that YORP provides a torque of a body using the asymmetry in its shape, Binary-YORP acts on the orbit of a satellite if it is tidally locked with its primary body, torqueing the entire system and changing the satellite's orbit. These two effects can operate simultaneously as tidal evolution, and potentially on much more rapid timescales (Jacobson & Scheeres 2011). The timescales are potentially so rapid that they could outpace formation mechanisms and lead to depletion of the binary population, which seems unlikely in light of the large fraction of the population having satellites. Thus, an alternative tidal formulation was devised that scales directly as the BYORP orbit evolution, so that the observed population would actually be in an equilibrium between tidal torques pushing the satellite outward and BYORP driving the satellite inward (Jacobson & Scheeres 2011). The critical tidal parameters of tidal dissipation $Q$ and rigidity by way of the tidal Love number $k_2$, would scale as $Q/k_2=6\times10^5$ ($R$/1km), whereas the prior rubble-pile tidal model had the scaling as $Q/k_2\geq10^7$ (1km/$R$). The ambiguities between the orbit and spin torques caused by tides and thermal effects may only be resolved through long timescale observations of binary systems or smoking-gun detections of the BYORP effect in action.

### 3.5 Cohesive strength

A well-known and glaring assumption in most early modeling of rubble pile dynamics was the lack of any tensile strength or physical bonding between individual constituent pieces. As pointed out by numerous authors tidal disruption models were both greatly simplified and scale in simple ways when this can be ignored, and when it was ignored models found utterly reasonable outcomes in terms of derived densities and morphologies for SL9 and also in terms of the expected rates of disruptions relative to the observed crater chains on the Galilean satellites (Schenk & Melosh 1993). Furthermore, calculations show that incredibly small amounts of cohesive strength can dramatically change outcomes - where ~3 Pascals was the maximum allowed cohesion in the best fit models for SL9's disruption (Asphaug & Benz 1994,1996, Holsapple & Michel 2006).

What would be the source of cohesion in a rubble pile? If the primary "grain" size is related to the 200-300m size regime where rapid spin rates start to be observed, then one could assume that this is a fundamental building block size (Walsh & Richardson 2006). Simple dry cohesive bonding by way of Van Der Waals force increases with decreasing grain sizes and would have no affect on such large rocks. However, images of asteroid Itokawa and inferences from thermal inertia data find that surfaces of rubble piles are covered with a wide range of much smaller grain sizes (Fujiwara et al. 2006, Delbo et al. 2015). If fine grains are also abundant in the interior of rubble piles then they could act as a sort of concrete with which to bind the larger constituent pieces of the rubble pile (Sánchez & Scheeres 2014). As the cohesion increases so do the spin limits for small asteroids in particular, with those ~km and smaller potentially experiencing faster allowable spin rates - such that the spin limit (Figure 2) would show a gradual increase in allowed spin rate at a large size depending on the cohesion limits. For example, 3 kPa of cohesion would permit 1km bodies to rotate faster than a 1hr period - which is not observed. Rather 100 Pa provides a comfortable envelope around the current set of observations (Sánchez & Scheeres 2014).

In order to make a strong case that a given body demands cohesion to explain its shape and spin requires accurate data on all of density, axis ratios and spin period. While mass is typically the hardest to acquire, it has also be backed out from the observed orbital drift of an asteroid due to the Yarkovsky effect when combined with estimates of the bodies' thermal inertia derived from thermal modeling of thermal emission (Chesley et al. 2014). This dataset has all of the information required to make this analysis, and Rozitis et al. (2014) found that for asteroid 1950 DA, its rapid spin rate of 2.1216 hr, density of 1.7±0.7 grams cm$^{-3}$ could not be explained with simply gravity and friction alone. The Drucker-Prager model for internal strength required derived $64^{+12}_{-20}$ Pa of cohesive strength required for it to maintain its shape. Finite element models attain similar required values of cohesion, 44-76 Pa, and describe a possible failure mode on the interior of the body resulting in equatorial bulging from the inside-out, rather than from mass movement from higher latitudes (Hirabayashi et al. 2015). The body also has remarkable thermal properties, evidenced by remarkably low Thermal Inertia, $24^{+20}_{-14}$ J m$^{-2}$ K$^{-1}$ s$^{-0.5}$, that strongly suggesting a surface covering of very small grains (the relationship between thermal response and surface properties is discussed in the next Section). This agrees in broad strokes with the description of cohesive binding of rubble piles by combined action of the smallest grains (Sánchez & Scheeres 2014), and with possible cohesion limits derived for a main belt comet that was observed just after rotational disruption (Hirabayashi et al. 2014).

# 4.0 Surface properties

What about the nature of a rubble pile can be studied or investigated by way of its surface? A handful of asteroids have been visited and studied up close, but not all of the surface geophysics is applicable to actual rubble piles and their high-porosity interiors. It is clear that even sub-km sized rubble piles have small particle surface covering - a "regolith" - that reveals numerous interesting geologic effects dominated by the local gravity (Murdoch et al. 2015). But, a rubble-pile itself consists of loose and unconsolidated particles and, by some definitions of regolith, could simply consist entirely of regolith-like material with no clear delineation between interior and surface materials. We adopt the definition of regolith as the "loose unconsolidated material that comprises the upper portions of the asteroid" (as defined by Robinson et al. 2002), which makes no real strict limitations on where the regolith turns into the subsurface or internals of a rubble pile. Fundamentally, the surface is what is observed astronomically, and what a spacecraft is interacting with, and that is what we seek to understand.

## 4.1 Nature of a rubble pile's regolith

The combined study of small asteroids' surfaces suggest that most are covered with some sort of relatively fine-grained material - called "regolith". This is not obvious, especially for small asteroids and comets, which were once assumed to simply be monolithic rocks with no surface covering. A regolith was inferred over time due to studies of brecciated meteorite textures, remote observations of asteroids and eventually spacecraft visits (Housen et al. 1982, Sullivan et al. 2002, Robinson et al. 2002, Fujiwara et al. 2006, Biele et al. 2015). The fundamental impediment for regolith generation and retention is the very low escape velocities of small asteroids (~1m/s for a 1km asteroid and increasing roughly linearly with diameter). Here, to zeroth-order, the collision speeds are so high in the main asteroid belt (averaging around 5km/s) compared to the relatively low asteroid escape velocities, that a large portion of the impact ejecta will simply escape the target asteroid and not contribute to its regolith (Chapman 1976, Housen et al. 1982). This logic led to the notion that deeper and finer regolith coverings should scale with size and be found on only the largest asteroids (Housen et al. 1982), and the accumulation of data on the thermal response of asteroid's regolith largely support this conclusion.

Spacecraft have now visited asteroids of many shapes and sizes, from the largest, Ceres, down to Itokawa, finding some sort of regolith on all of them. These visits have validated many of the ground-based techniques and inferences made, but also established that small rubble pile asteroids, despite having less fine-grained regolith than larger bodies, still have interesting and complex surface geology. Before *in situ* data is available for an asteroid significant remote sensing capabilities are available to constrain the nature of its surface.

### 4.1.1 Thermal inertia and thermal fatigue

Regolith with different size grains have different thermal properties where large boulders and rocks retain daytime heat and re-radiate that heat slowly during nighttime. Meanwhile very fine grains quickly release any stored heat and provide a significantly different surface temperature profile over time. Observations in the thermal infrared wavelengths (where the wavelength correspond to the thermal emission from the surface, which depends on temperature and therefore distance from the Sun) can provide the data to model this behavior and constrain numerous properties about the surface (Delbó et al. 2015).

The thermal inertia of a body reflects the material heat capacity and the thermal conduction into and out of the subsoil, or deeper layers of the regolith. As thermal inertia goes towards zero the surface temperature increasingly matches the incoming solar flux leading to peak surface temperatures at noon with near-zero surface temperature across the nightside surface. As thermal inertia increases, this diurnal surface temperature curve get smoothed out. Absolutely critical to this process is the huge difference in conduction between large solid rocks and very fine-grained powdery lunar regolith, which can differ by three orders of magnitude. Therefore thermal inertia measurements relative to the Moon provide insight into typical grain sizes on an asteroid's surface.

A simple thermal modeling technique used when there is minimal knowledge of an asteroid's shape or spin is the near-Earth Asteroid Thermal Model (NEATM) (Harris 1998). NEATM necessarily makes assumptions about an asteroid's shape (spherical) to estimate an asteroid's diameter and albedo (Harris 1998), and encodes information about thermal inertia and surface roughness in a "beaming parameter", that can sometimes be used for comparisons of large numbers of objects (Delbó et al. 2007). More direct measurements of thermal inertia require full thermophysical models that rely on shape models of the target body and knowledge of its spin state. With this data in hand calculations for each surface facet can be made resulting in much stricter constraints on thermal inertia and surface roughness across an asteroid (Delbó et al. 2015).

The measured values of thermal inertia for asteroids is found to be size dependent (see Figure 12), where the largest asteroids have values approaching that for the Moon (~50 in these SI units, $J\ m^{-2}\ s^{-½}\ K^{-1}$), rapidly decreasing at smaller sizes. Rubble pile Itokawa provides a valuable marker with a thermal inertia of $700 \pm 200\ J\ m^{-2}\ s^{-½}\ K^{-1}$ (Mueller 2007, Müller et al. 2014), in line with the imagery showing alternatively very rocky and boulder dominated terrain across much of its surface. Furthermore, relying on the physical foundation of thermal inertia being built on conduction between touching grains formalisms for turning a measured thermal inertia value into a grain size, or average grain size, have been built and can be compared to these measured values (Gundlach and Blum 2013). The asteroid Itokawa demonstrates the challenges in interpreting these calculations, as the surface shows very wide variations and no singular average grain size can fully describe it - despite the calculation estimating ~22 mm. Using this to make a relative comparison, for the asteroid Bennu, with its lower thermal inertia ($310 \pm 70$), suggests at the very least a surface with *more* smaller grains throughout (Emery et al. 2014).

The same diurnal temperature cycles that allow astronomers to probe an asteroid's thermal properties also stress the grains and rocks that are being heated and cooled many times a day for millions or billions of years. This regular and constant thermal cycling has been explored in the lab and extrapolated in models to suggest that it could fracture and break down boulders and rocks and play a role in forming the fine grains found on asteroid surfaces (Delbo et al. 2014). Compared to an impact generation process for regolith production, which depends on the relationship between impact ejecta speeds and the bodies size and escape speed, thermal fatigue and cracking is size independent. This process may also be related to the loss of small bodies that have low perihelion - a notable absence discovered in recent modeling work – that could be thermal fatigue working in very destructive ways at much higher temperatures (Granvik et al. 2016)

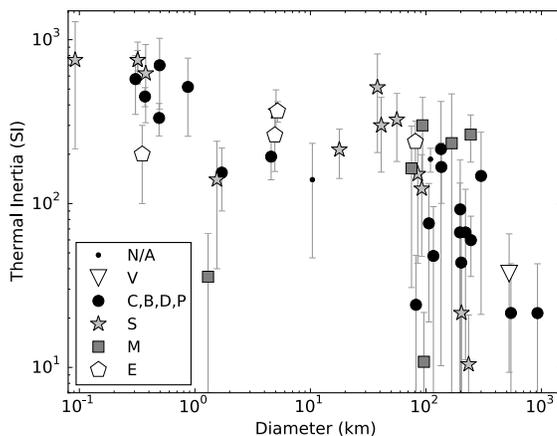

[**Figure 13:** Thermal inertia values for asteroids in units of $J\ m^{-2}\ K^{-1}\ s^{-0.5}$ plotted against their diameter in kilometers. The data are broken down into various asteroid taxonomy, but the trend stands for all for all them with higher thermal inertia at smaller sizes. Plot taken from Delbo et al. (2015).**]**

### 4.1.2 Radar reflectivity and roughness

Radar observations measure echo power from an object in both the time delay and doppler frequency, but they also include circular polarization recorded in terms of the opposite sense (OC) as transmitted and waves received in the same sense as transmitted (SC). In perfectly smooth scattering (relative to the lengthscale of the radar wavelength) the OC would dominate the return signal, giving a polarization ratio SC/OC=0. A surface that could completely randomize the reflected signal would approach SC/OC=1. For a given wavelength of observation the SC echos increase, and so does the SC/OC ratio, with increased surface roughness in length scales similar to that of the transmission wavelength (Ostro et al. 2002).

This method of observation has been "ground-truthed" by data from Itokawa (Ostro et al. 2004, Miyamoto et al. 2007). Itokawa was observed in both 3.5 and 12.6cm wavelength radar, finding higher polarization at 3.5 than at 12.6cm, suggesting a changing surface texture at size scales between these two values (Ostro et al. 2004, Nolan et al. 2013). Meanwhile Bennu has a lower polarization ratios than Itokawa and similar values at both 3.5cm and 12.6, suggesting that Bennu has a larger population of few-cm size grains relative to Itokawa and characteristic roughness scales smaller than 3.5cm (Nolan et al. 2013). The ground-truth for Itokawa is not straightforward, as the surface is far from homogenous, with some very rocky and boulder-laden terrains and other ponds of very fine grains, but further ground-truths of these techniques should be obtained with the OSIRIS-REx visit to the NEA Bennu.

### 4.2 Observed Surface Geology

Itokawa is one of the best studied asteroids and the only rubble pile to be visited by spacecraft and therefore necessarily dominates discussion of surface geology on rubble pile asteroids. It showed that gravity is important even on very small asteroids, with gravitational slopes and potential lows being correlated with surface features. The absence of a large crater population or any large linear features combined with signs of regolith migration point to seismic shaking as an important energy source for surface modification (see Figure 1 & 14: Miyamoto et al. 2007; Barnouin-Jha et al. 2008).

Itokawa had both rough and smoother terrains nominally defined by topography, with rough highlands and smooth lowlands (Saito et al. 2006, Miyamoto et al. 2007, Barnouin-Jha et al. 2008). The rougher highlands display surface roughness on scales of 2-4m, indicating a covering of meter-size boulders, and appear devoid of particles smaller than 1 cm (Barnouin-Jha et al. 2008, Miyamoto et al. 2007). Within the highlands there are a handful of smoother regions associated with localized depressions (Barnouin-Jha et al. 2008). The smoother lowlands are flat to sub-meter scales and essentially devoid of boulders (Barnouin-Jha et al. 2008), with some indications of possible crater-like depressions (Hirata et al. 2009). The roughness increases towards the edges of the smooth regions, which has led to the interpretation that the finer grains are migrating into the low-elevation areas and covering up the otherwise rocky terrain (Miyamoto et al. 2007, Barnouin-Jha et al. 2008). The boundary between the rough and smooth terrains shows signs of regolith movement and migration, with imbricated boulders aligned with local gravity slopes, pileups of small grains behind larger boulders and strong alignments of large irregularly shaped boulders (Miyamoto et al. 2007). The presence of rounded boulders similarly points to active geologic processes both moving and processing large surface particles (Marshall and Rizk 2015, Connolly et al. 2015.). The implied direction of these signs of regolith migration are in line with the local gravitational slope and demand a source of energy for their movement.

  The small size and surface gravity of Itokawa, along with the global differences in terrain associated with potential lows and highs, point to seismic shaking as the key geologic driver on its surface (Miyamoto et al. 2007, Richardson et al. 2005). The combined action of small impacts result in effective gravel fluidization as impactors on cm-scales can impart enough energy to rival the surface escape speed regionally (Cheng et al. 2002). A question for this effect dominating on rubble piles is how widely the energy can be transported through a non-coherent and high-porosity interior. Miyamoto et al. (2007) suggested that seismic shaking could drive finer grains into the interior and subsequently change the energy attenuation of its non-monolithic rubble-pile interior. Both upcoming space missions to rubble piles, Hayabusa2 and OSIRIS-REx, will have the opportunity to look for further signs of these processes while also contrasting the surface geology between different taxonomic types of asteroids.

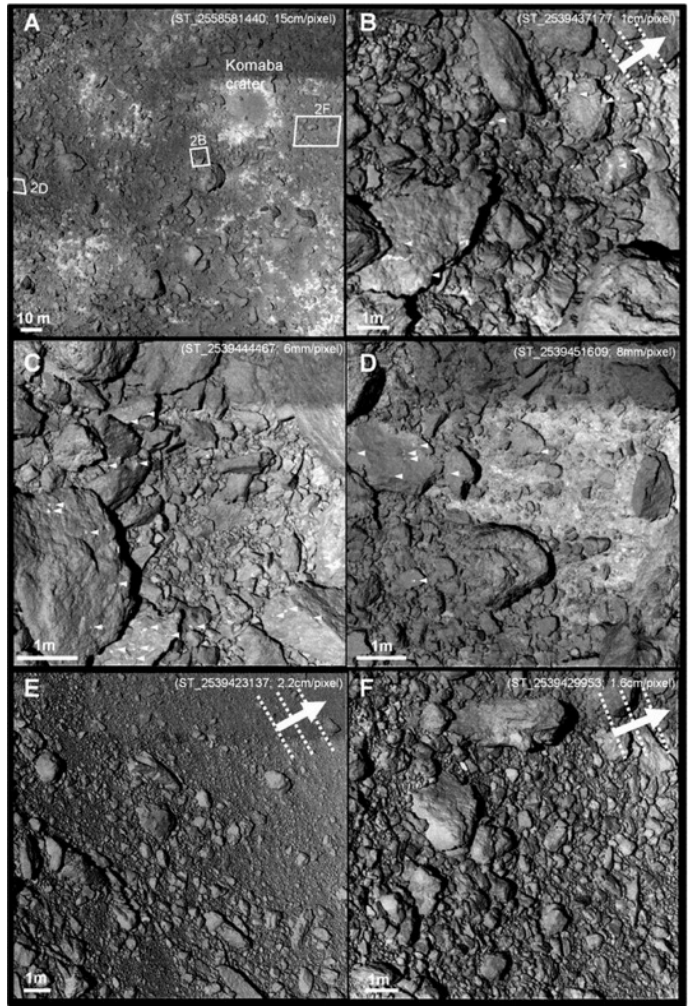

[**Figure 14:** The surface of Itokawa as imaged from JAXA's Hayabusa spacecraft taken directly from Miyamoto et al. (2007). A) This is the terrain around the Komaba crater at the resolution of 15cm/pixel, and the outlined boxes show the field of view for panes B, D and F; B) Higher resolution image showing the organization of grains following the indicated lines and suggestive of flow in the direction of the arrow, C) 6mm/pixel resolution shows large boulders piled on top of smaller regolith and the arrows indicate bright marking on some of the boulders, D) more rough terrain and strong albedo contrast, E) pile up of small grains is found on the uphill side of larger rocks in this terrain between rough highlands the smooth Muses C region and, F) in another area between smooth and rough terrain there are signs of imbrication.]

# 5.0 Exploration

Space exploration has advanced to the point where spacecraft regularly interact with the surfaces of planets (Mars and Venus), large moons (the Moon and Titan), asteroids (Eros and Itokawa) and also comets (9P/Tempel 1 and 67P/Churyumov-Gerasimenko). Of these Itokawa clearly fits the profile of being a rubble pile, partly due to its striking appearance. The interaction with the surface of Itokawa was brief, but the close-up imagery has provided a wealth of science. The upcoming space missions will do more and record more during their attempts to sample the surfaces of small rubble pile asteroids Ryugu (target of JAXA's Hayabusa2) and Bennu (target of NASA's OSIRIS-REx), and hopefully set the stage for even more ambitious efforts in the future. Both of these missions aim to return samples from the surface of their respective targets utilizing very different collection techniques.

While the combined knowledge of rubble pile asteroids can provide reference points for planning and provide tests for interpreting our observational and analysis techniques, it is also useful to ask what absolutely new things should be expected to come from these gigantic efforts. Both missions will survey their targets in an array of wavelengths, but both will also interact with their targets in completely new ways, which should provide important data about their surfaces and sub-surfaces.

## 5.1 The Hayabusa2 cratering experiment

JAXA's Hayabusa2 mission will undertake a spectacular experiment designed to reveal the subsurface of its target asteroid Ryugu. The spacecraft, launched in December, 2014, carries a kinetic impactor, the small carry-on impactor (SCI), which is designed to accelerate a 2kg mass to 2km/s in order to make a sizeable crater on the surface of Ryugu (Saiki et al. 2017, Arakawa et al. 2017). Due to the violence of the explosion needed to accelerate such and impactor, and the possible debris launched during the formation of the crater on the surface of the asteroid, the main spacecraft will be moved out of view and a deployable camera will instead witness the event (Ishibashi et al. 2017).

The resulting crater is intended to reveal fresh material for later sampling, but also for remote sensing of this uncovered subsurface, unweathered, material to compare with the rest of the asteroid surface. The crater itself, its size and morphology, provides information about the near and subsurface of the target - whereby a singular large block being impacted provides strength-dominated cratering resulting in a crater up to 10m across (Arakawa et al. 2017). Meanwhile an impact in a boulder field would still expend much of the energy into breaking individual rocks, but respond with a much smaller (as little as 1/10th the size) crater. Fluffy and porous layers on top of either of these pushes some of the response into the gravity-regime and keep the final crater relatively small as well. Meanwhile the deployed camera will view the ejecta curtain and should constrain the ejecta velocity distribution (Arakawa et al. 2017).

This impact should provide an instrument to test some of the large ideas around regolith migration on small bodies by actively driving a seismic shaking experiment. One of the primary inferred actions of seismic shaking has been the degradation of small craters and crater walls that could be observable (Richardson et al. 2005, Robinson et al. 2002, Michel et al. 2009), but the transmission of a seismic wave through the regolith layer (Yasui et al. 2015) and throughout the subsurface structure should provide some information on the internal structure of Ryugu.

## 5.2 The OSIRIS-REx surface interaction

Bennu, the target of NASA's OSIRIS-REx mission, is very well characterized with astronomincal techniques, with a high resolution shape model and surface roughness characterization from radar observations (Nolan et al. 2013), thermal inertia from thermal infrared observations (Emery et al. 2014), and a density determined from analysis of orbital change due to the Yarkovsky effect (Chesley et al. 2014). Much of the discussion above made comparisons between Bennu and Itokawa based on this data, typically pointing toward expectations of an asteroid with characteristic surface roughness at the cm-scales and average grain sizes below cm-sizes (Emery et al. 2014).

The OSIRIS-REx surface interaction with its target Bennu relies on a new and novel sampling approach that minimizes contact time with the surface by utilizing a touch-and-go system (Lauretta et al. 2017). Following the survey of its target OSIRIS-REx will do a series of navigation maneuvers to move from a 1-km orbit, onto a trajectory to intercept the asteroid, another series of maneuvers to match the lateral motion of the target sample site due to the asteroid's rotation, and then finally contacting the surface at 10cm/s with minimal lateral velocity (±2cm/s) (Lauretta et al. 2017). To combat any cohesive bonding or induration in the sampling area pyro valves will open and release high-purity nitrogen gas after contact with the surface is determined. The gas will also redirect the fluidized material into the sampling collector that is extended on a 2.8m arm extended from the main spacecraft's body. The sampling head is an annulus, ringed by a screen to capture material that is flowing through the device driven by the released gas. The sampling head also has 24 contact pads made of stainless steel meant "grab" very small particles (Lauretta et al. 2017).

This sampling attempt is a controlled science experiment on its own, as the spacecraft will impact the surface of Bennu at ~10cm/s with a known mass and at a known spot on the surface studied that has been surveyed in numerous wavelengths. While the returned sample will provide very tangible data on the material at the point of contact, the dynamics of the spacecraft during sample will probe the upper layers of the surface of Bennu. Spacecraft telemetry will record accelerations on the main spacecraft with high frequency so that the force imparted on the spacecraft during contact will be recorded. The spacecraft is dynamically separated from the asteroid's surface by a constant-force spring, and there is a maximum time allowed before the backaway burn will commence, so the total interaction is quite short, but while the spring will obscure some of the force imparted by the asteroid there will likely be a record of the initial contact and penetration (Lauretta et al. 2017). Force profiles during contact/impact can be used to characterize the material properties of the surface (Goldman & Umbanhowar 2008) and can hopefully be used to detect the nature of the regolith – whether there are signs of stratification or layering at the surface, as inferred at the surface of comet 67P by the dynamics of the Philae lander (Schrapler et al. 2015, Biele et al. 2015).

# 6.0 Conclusion

Referring to small asteroids as "rubble piles" works. It describes all the lines of evidence and they support this generic moniker. From the thousands of collected lightcurves down to the rubbly and size-sorted surface of Itokawa, no data suggests that this is an unfair assessment of their qualitative nature. Furthermore, quantitative assessments of resistance to re-shaping, possible cohesive bonding, and comparisons against shear properties of standard terrestrial granular media really support them being just gravitational bound piles of rocks.

But, with just one of these objects studied up close, and two more in the sights of spacecraft, these ideas will all be tested again. Tripling the number of bodies for which surface geology has been studied and surface touched should enhance the data presented here and hopefully surprise the scientific community all over again.

**SUMMARY POINTS**:

**FUTURE ISSUES**:

1. Detecting and understanding the Binary-YORP effect will be essential to understanding tidal evolution of rubble pile with satellites and learning more about their tidal dissipation.

2. Bennu has the ubiquitous top-shape suggestive of past spin up and reshaping, but it has a moderate spin rate and no satellite. Will it show signs of re-shaping, and if so, will it point towards internal failure or surface shedding and landslides?

3. JAXA's Hayabusa2 spacecraft will make a crater with its small carry-on impactor – will the crater formation process reveal much about the interior structure of Ryugu or will evidence of seismic shaking appear globally from the impact?

4. Are carbonaceous rubble pile similar, in bulk, to Itokawa or will the primitive asteroids Ryugu and Bennu, both set to be visited by spacecraft show different global properites?

5. What is the role of cohesion in rubble pile behavior and evolution? Will more novel studies constrain these values from astronomical data or will space mission visits shed more light by way of surface geology?

**ACKNOWLEDGEMENTS**:

K.J.W would like to acknowledge support and inspiration from NASA OSIRIS-REx asteroid sample return mission.

**REFERENCE ANNOTATION** (15 word maximums):

1. Asphaug & Benz 1996 : This study used tidal breakup morphology to estimate SL9's density

2. Bottke et al. 1999: An amazing match between a tidal disruption simulation and radar-observed NEA.

3. Chapman et al. 1978: First use of the term "rubble pile" in the literature

4. Ćuk & Burns 2005: First derivation of the Binary-YORP effect

5. DeMeo & Carry 2014: Modern overview of asteroid physical and taxonomic distribution in the Main Belt and beyond

6. Gundlach & Blum 2013: Techniques to turn an observed thermal inertia into an average grain size.

7. Holsapple 2001: Important derivation of allowed shapes for cohesionless asteroids.

8. Margot et al. 2002: First radar discovered binary asteroid

9. Michel et al. 2001: End-to-end model of asteroid disruption, dispersal and reaccumulation of fragments

10. Yasui et al. 2015: Important study on seismic shaking

**SIDEBAR (200 words max – 183 present):**

The YORP-effect relies on the asymmetries of an asteroid's shape to produce a torque around its spin axis when it reflects or re-emits solar radiation (Rubincam 2000, Bottke et al. 2006). The action of the YORP effect is directly observed in a handful of objects that are observed to be actively spinning up (Taylor et al. 2007, Kaasalainen et al. 2007) and long-term evolution is found in populations of asteroids that are driven into spin-orbit resonances in the Main Belt ("Slivan States" - Slivan et al. 2002; Vokrouhlický et al. 2003) and in the way that asteroid families drift apart over time (Vokrouhlický et al. 2006, Nesvorný et al. 2015). The YORP-effect can also alter obliquities, effectively pushing asteroids towards 0 or 180deg obliquities (Vokrouhlicky & Čapek 2002, Čapek and Vokrouhlický 2004). Whether or not a specific asteroid is likely to be spinning up or spinning down is not-trivial to calculate, even for a well-studied body with a known shape, and some studies suggest that even small scale surface artifacts like boulders or very small craters can change how a body responds (Statler 2009).

**FIGURES**: